\documentclass[12pt]{iopart}
\usepackage{amssymb}
\usepackage{graphicx}
\usepackage{graphics}
\usepackage{epsfig}
\begin{document}

\title[Detecting order and chaos by the SALI method]
{Detecting order and chaos in Hamiltonian systems by the SALI method}

\author{Ch Skokos\dag\footnote[2]{Corresponding author:
Ch Skokos},  Ch Antonopoulos\dag,
T C Bountis\dag \, and M N Vrahatis\S}

\address{\dag\ Department of Mathematics, Division of Applied
Analysis and Center for Research and Applications of Nonlinear
Systems (CRANS), University of Patras, GR-26500 Patras, Greece}

\address{\S\ Department of Mathematics and University of Patras Artificial
Intelligence Research Center (UPAIRC), University of Patras,
GR-26110 Patras, Greece}

\ead{hskokos@cc.uoa.gr}

\begin{abstract}
We use the Smaller Alignment Index (SALI) to distinguish rapidly
and with certainty between ordered and chaotic motion in
Hamiltonian flows. This distinction is based on the different
behavior of the SALI for the two cases: the index fluctuates
around non--zero values for ordered orbits, while it tends rapidly
to zero for chaotic orbits. We present a detailed study of SALI's
behavior for chaotic orbits and show that in this case the SALI
exponentially converges  to zero, following a time rate depending
on the difference of the two largest Lyapunov exponents
$\sigma_1$, $\sigma_2$ i.e. $\mbox {SALI} \propto
e^{-(\sigma_1-\sigma_2)t}$. Exploiting the advantages of the SALI
method, we demonstrate how one can rapidly identify even tiny
regions of order or chaos in the phase space of Hamiltonian
systems of 2 and 3 degrees of freedom.
\end{abstract}


\submitto{\JPA}

\maketitle

\section{Introduction}
\label{intro}
Knowing whether the orbits of a dynamical system are ordered or
chaotic is fundamental for the understanding of the behavior of
the system. In the dissipative case, this distinction is easily
made as both types of motion are attracting. In conservative
systems, however, distinguishing between order and chaos is often
a delicate issue (e.g.\ when the chaotic or ordered regions are
small) especially in systems with many degrees of freedom where
one cannot easily  visualize the dynamics. For this reason it is
of great importance to have quantities that determine if an orbit
is ordered or chaotic, independent of the dimension of its phase
space.

The well--known and commonly used method for this purpose is the
evaluation of the maximal Lyapunov Characteristic Exponent (LCE)
$\sigma_1$. If $\sigma_1 > 0$ the orbit is chaotic. Benettin et
al.\ \cite{BGGS80a} studied theoretically the problem of the
computation of all LCEs and proposed in \cite{BGGS80b} an
algorithm for their numerical computation. In particular,
$\sigma_1$ is computed as the limit for $t \rightarrow \infty$ of
the quantity
\begin{equation}
L_t=\frac{1}{t}\, \ln  \frac{|\vec{w}(t)|}{|\vec{w}(0)|}\, ,\,
\mbox{i.e.}\,\, \sigma_1 = \lim_{t\rightarrow \infty} L_t \, ,
\label{eq:lyap}
\end{equation}
where $\vec{w}(0)$, $\vec{w}(t)$ are  deviation vectors from a
given orbit, at times $t=0$ and $t>0$ respectively. The time
evolution of $\vec{w}$ is given by solving the so--called {\it
variational equations} (see Sec. \ref{Behavior}). Generally, for
almost all choices of initial deviations $\vec{w}(0)$, the limit
for $t \rightarrow \infty$ of Eq.~(\ref{eq:lyap}) gives the same
$\sigma_1$.

In practice, of course, since the exponential growth of
$\vec{w}(t)$ occurs for short time intervals, one stops the
evolution of $\vec{w}(t)$ after some time $T_1$, records the
computed $L_{T_1}$, normalize vector $\vec{w}(t)$
and repeats the calculation for another time
interval $T_2$, etc. obtaining finally $\sigma_1$ as an average
over many $T_i$, $i=1,2,\ldots,N$ as
\begin{displaymath}
\sigma_1 = \frac{1}{N} \sum_{i=1}^{N} L_{T_i}
\end{displaymath}
The basic problem of the computation of $\sigma_1$  is that, after
every $T_i$, the calculation starts from the beginning and may
yield an altogether different $L_{T_i}$ than the $T_{(i-1)}$
interval. Thus, since $\sigma_1$ is influenced by the whole
evolution of $\vec{w}(0)$, the time needed for $L_t$ (or the
$L_{T_i}$) to converge is not known a priori and may become
extremely long. This makes it often difficult to tell whether
$\sigma_1$ finally tends to a positive value (chaos) or converges
to zero (order).

In recent years, several methods have been introduced which try to
avoid this problem by studying  the evolution of deviation
vectors, some of which are briefly discussed in Sec.\
\ref{Compare}. In the present paper, we focus our attention on the
method of the Smaller Alignment Index (SALI)~\cite{SALI},
performing a systematic study of its behavior in the case of
autonomous Hamiltonian systems with 2 (2D) and 3 (3D) degrees of
freedom. This method has been applied successfully to several
2--dimensional (2d) and multidimensional maps \cite{SALI}, where
SALI was found to converge rapidly to zero for chaotic orbits,
while it exhibits small fluctuations around non--zero values for
ordered orbits. It is exactly this ``opposite" behavior of the
SALI which makes it an ideal indicator of chaoticity: Unlike the
maximal LCE, it does not start at every step a new calculation of
the deviation vectors, but takes into account information about
their convergence on the unstable manifold from all the previous
steps. The method has already been used successfully as a chaos
detection tool in some specific Hamiltonian systems
\cite{SABV03c,PBS04,S03,SESS04,VKS02,KVC04}, although some authors
\cite{VKS02,KVC04} use different names for the SALI.

The paper is organized as follows: In Sec.\ \ref{Ap} we recall the
definition of the SALI and present  results distinguishing between
ordered and chaotic motion in 2 and 3--degrees of freedom (2D and
3D) Hamiltonians, comparing also the efficiency of the SALI with
the computation of $\sigma_1$. In Sec.\ \ref{Behavior} we explain
the behavior of the SALI  for ordered and chaotic orbits, showing
that in the latter case SALI converges exponentially to zero
following a rate which depends on the difference of the two
largest Lyapunov exponents $\sigma_1$ and $\sigma_2$. In Sec.\
\ref{Discussion} we demonstrate the ability of the method to
reveal the detailed structure of the dynamics in the phase space.
In Sec.\ \ref{Compare} we compare the SALI method with some other
known methods of chaos detection and in Sec.\ \ref{Summary} we
summarize our results.

\section{Application of the SALI in Hamiltonian systems}
\label{Ap}

The basic idea behind the success of the SALI method \cite{SALI}
is the introduction of a simple quantity that clearly indicates if
a deviation vector is aligned with the direction of the
eigenvector which corresponds to the maximal LCE. In general, any
two randomly chosen initial deviation vectors $\vec{w}_1 (0)$,
$\vec{w}_2 (0)$ will become aligned with the most unstable
direction and the angle between them
will rapidly tend to zero \cite{BGGS80b}. Thus, we check
if the two vectors have the same direction in phase space, which
is equivalent to the computation of the above--mentioned angle.

More specifically, we follow simultaneously the time evolution of
an orbit with initial condition $\vec{x} (0)$ and two deviation
vectors with initial conditions $\vec{w}_1 (0)$, $\vec{w}_2 (0)$.
As we are only interested in the directions of these two vectors
we normalize them, at every time step, keeping their norm equal to
1. This controls  the exponential increase of the norm of the
vectors and avoids overflow problems. Since, in the case of
chaotic orbits the normalized vectors point to the same direction
and become equal or opposite in sign, the minimum of the norms of
their sum ({\it antiparallel alignment index}) or difference ({\it
parallel alignment index}) tends to zero. So the SALI is defined
as:
\begin{equation}
\mbox{SALI}(t)= \min \left\{ \left\|
\frac{\vec{w}_1(t)}{\|\vec{w}_1(t)\|}+
\frac{\vec{w}_2(t)}{\|\vec{w}_2(t)\|} \right\| , \left\|
\frac{\vec{w}_1(t)}{\|\vec{w}_1(t)\|}
-\frac{\vec{w}_2(t)}{\|\vec{w}_2(t)\|} \right\| \right\},
\label{eq:SALI}
\end{equation}
where $t$ is the  time and $\|\cdot\|$ denotes the Euclidean norm.
From the above definition is evident that $\mbox{SALI}(t) \in
[0,\sqrt{2}]$ and when $\mbox{SALI}=0$
the two normalized vectors have the same
direction, being equal or opposite.

In order to apply the SALI method to Hamiltonian systems, we shall
use here two simple examples with 2 and 3 degrees of freedom: the
well--known 2D H\'{e}non--Heiles system \cite{HH64}, having the
Hamiltonian function:
\begin{equation}
H_2 = \frac{1}{2} (p_x^2+p_y^2) + \frac{1}{2} (x^2+y^2) + x^2 y -
\frac{1}{3} y^3, \label{eq:2DHam}
\end{equation}
with equations of motion:
\begin{equation}
\ddot{x}=-x-2xy \, , \,\,\, \ddot{y}=-y-x^2+y^2\, ,
\label{eq:2DHameq}
\end{equation}
and the 3D Hamiltonian system:
\begin{equation}
H_3 = \frac{1}{2} (p_x^2+p_y^2+p_z^2) + \frac{1}{2}
(Ax^2+By^2+Cz^2)- \epsilon x z^2 - \eta y z^2, \label{eq:3DHam}
\end{equation}
with equations
\begin{equation}
\ddot{x}=-Ax+\epsilon z^2 \, , \,\,\, \ddot{y}=-By+\eta z^2 \, ,
\,\,\, \ddot{z}=-Cx + 2z(\epsilon x + \eta y)\, ,
\label{eq:3DHameq}
\end{equation}
studied in \cite{CM85,CB89}. We keep the parameters of the two
systems fixed at the energies $H_2=0.125$ and $H_3=0.00765$, with
$A=0.9$, $B=0.4$, $C=0.225$, $\epsilon=0.56$ and $\eta=0.2$.

A simple qualitative way of studying the dynamics of a Hamiltonian
system is by plotting the successive intersections of the orbits
with a Poincar\'{e} surface of section (PSS) \cite{LL92}. This
method has been extensively applied to 2D Hamiltonians, as in
these systems the PSS is a 2--dimensional plane. In 3D systems
however, the PSS is 4--dimensional and the behavior of the orbits
cannot be easily visualized. One way to overcome this problem is
to project the PSS to spaces with lower dimensions (see e.g.\
\cite{VBK96,VIB97}). However, even these projections are often
very complicated and difficult to interpret.

\begin{figure}
\centerline{\includegraphics[width=15 cm,height=7.5 cm]
{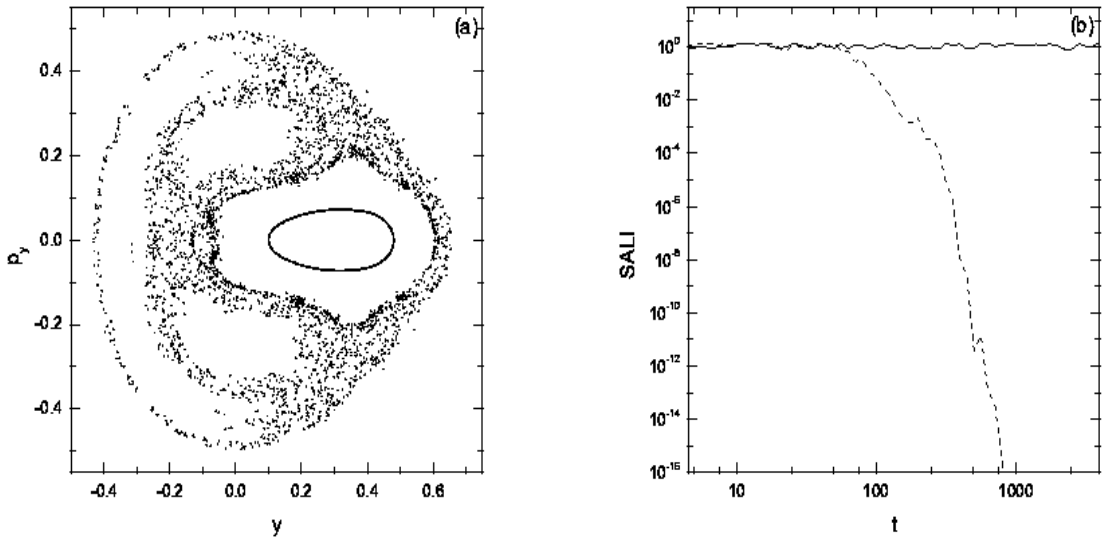}}
\caption{(a) The PSS of an ordered  and a chaotic
orbit  with initial conditions $x=0$, $y=0.1$, $p_x\simeq0.49058$,
$p_y=0$ and $x=0$, $y=-0.25$, $p_x\simeq0.42081$, $p_y=0$
respectively, for the H\'{e}non--Heiles system (\ref{eq:2DHam}).
The ordered orbit corresponds to a closed (solid) elliptic curve,
while the chaotic one is represented by the dots scattered over
the PSS. (b) The time evolution of the SALI for the two orbits of
panel (a) in log-log scale. The solid line corresponds to the
ordered orbit while the dashed line corresponds to the chaotic
orbit.} \label{fig:2Dexample}
\end{figure}

\begin{figure}
\centerline{\includegraphics[width=7.5 cm,height=7.5 cm]
{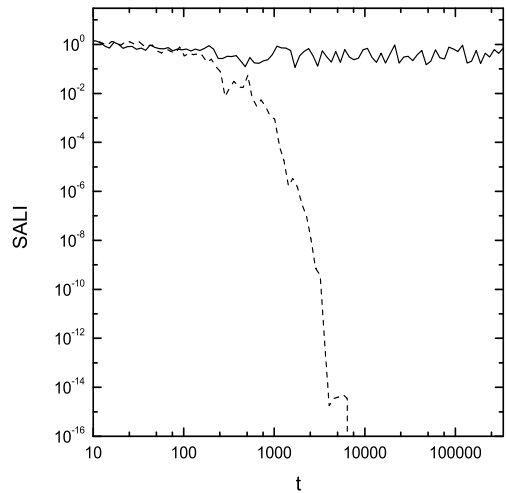}}
\caption{The time evolution of the SALI for an ordered
(solid line) and a chaotic orbit (dashed line) of the 3D
Hamiltonian (\ref{eq:3DHam}), with initial conditions $x=0.01054$,
$y=0.05060$, $z=0$, $p_x=0$, $p_y=0$, $p_z\simeq0.11906$ and
$x=-0.07310$, $y=0$, $z=0$, $p_x=0.07695$, $p_y=0$,
$p_z\simeq0.06760$ respectively.} \label{fig:3Dexample}
\end{figure}

In order to illustrate the behavior of the SALI in 2D and 3D
systems we first consider some representative ordered and chaotic
orbits. In Fig. \ref{fig:2Dexample}(a)
we plot the intersection points of an ordered  and a chaotic orbit
of Eqs. (\ref{eq:2DHameq}), with a PSS defined by $x=0$. The
points of the ordered orbit  lie on a torus and form a smooth
closed curve on the PSS. On the other hand, the points of the
chaotic orbit  appear randomly scattered. The time evolution of
the SALI for these two orbits is plotted in Fig.
\ref{fig:2Dexample}(b). In the case of the ordered orbit (solid
line) the SALI remains different from zero, while in the case of
the chaotic orbit (dashed line), after a small transient time, the
SALI falls abruptly to zero. At $t\approx 800$ the SALI becomes
zero as it has reached the limit of the accuracy of the computer
($10^{-16}$), which means that the two deviation vectors have the
same direction. Thus, after $t \approx 800$ the two normalized
vectors are represented by exactly the same numbers in the
computer and we can safely argue, that to this accuracy the orbit
is chaotic. Actually, we could conclude that the orbit is chaotic
even sooner, considering that the directions of the two vectors
{\em practically} coincide when the SALI reaches a small value,
e.g.\ $10^{-8}$ after some $400$ time units. Entirely analogous
behavior of the SALI distinguishes between ordered and chaotic
orbits of the 3D Hamiltonian (\ref{eq:3DHam}), as shown in Fig.
\ref{fig:3Dexample}.

\begin{figure}
\centerline{\includegraphics[width=15 cm,height=7.5 cm]
{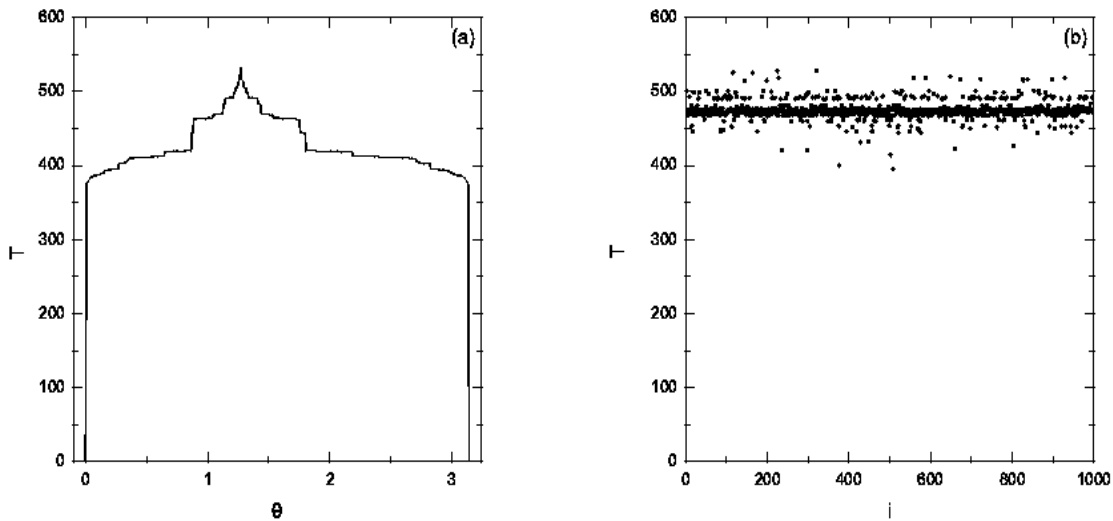}} \caption{The time $T$ needed for the SALI to become
less than $10^{-12}$ in the case of the chaotic orbit of Fig.\
\ref{fig:2Dexample} when we use as initial deviation vectors
$\vec{w}_1(0)=(0,1,0,0)$ and (a) $\vec{w}_2(0)=(0,\cos \theta, 0,
\sin \theta)$, and (b) $\vec{w}_2(0)$ being a random vector. $T$
is plotted as a function of $\theta$ in (a), and  as a function of
a counter $i$ of the randomly generated vectors  in (b).}
\label{fig:random}
\end{figure}

The initial deviation vectors, $\vec{w}=(dx, dy, dp_x, dp_y)$,
used for both orbits of Fig.\ \ref{fig:2Dexample} are
$\vec{w}_1(0)=(1,0,0,0)$, $\vec{w}_2(0)=(0,0,1,0)$, but in general
any other initial choice leads to similar behavior of the SALI.
The validity of the above statement is supported by the following
computations. Focusing our attention on the more interesting case
of chaotic motion we study the chaotic orbit of Fig.\
\ref{fig:2Dexample} by fixing one of the initial deviation
vectors, keeping it e.\ g.\ $\vec{w}_1(0)=(0,1,0,0)$ and varying
the second one $\vec{w}_2(0)$. For every different pair of initial
vectors $\vec{w}_1(0)$, $\vec{w}_2(0)$ we compute the time $T$
needed for the SALI to become smaller than a very small value e.\
g.\ $10^{-12}$ and check if the value of $T$ depends on the
particular choice of initial deviation vectors. We choose
$\vec{w}_2(0)$ in two different ways. Firstly we consider the two
initial vectors to be on the PSS of Fig.\ \ref{fig:2Dexample}
having an angle $\theta$ between them so that
$\vec{w}_2(0)=(0,\cos \theta, 0, \sin \theta)$. In Fig.\
\ref{fig:random}(a)
we plot $T$ as a function of $\theta$ for $\theta \in [0, \pi]$.
As expected $T=0$ for $\theta=0$ and $\theta=\pi$ since the two
vectors are initially aligned. The maximum value of $T$ for
$\theta \approx 0.405 \pi$ corresponds to the case of
$\vec{w}_2(0)$ being almost perpendicular to the unstable manifold
which passes near the initial condition of the orbit. In this case
the component of $\vec{w}_2(0)$ along the unstable direction is
almost zero and thus, the time needed for the vector to develop a
significant component along this direction, which will eventually
lead it to align with the other deviation vector, is maximized. We
remark that for all $\theta \in (0, \pi)$ $T$ does not change
significantly, as it practically varies between 400 and 500 time
units. As there is no reason for $\vec{w}_1(0)$, $\vec{w}_2(0)$ to
be on the PSS the second test we perform is to compute $T$ for
1000  $\vec{w}_2(0)$ whose coordinates are randomly generated
numbers (Fig.\ \ref{fig:random}b). From the results of Fig.\
\ref{fig:random} we see that $T$ practically  does not depend on
choice of the initial deviation vectors.

\begin{figure}
\centerline{\includegraphics[width=15 cm,height=7.5 cm]
{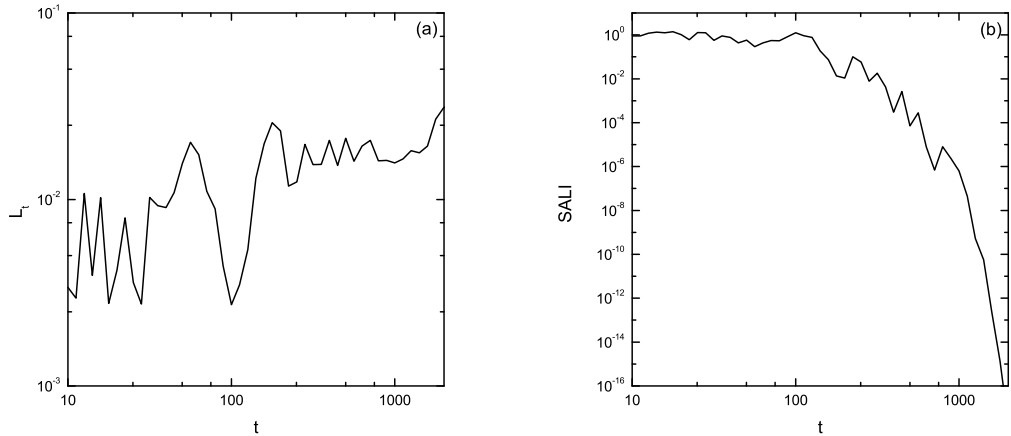}} \caption{The time evolution of (a) the $L_t$  and (b)
the SALI for the chaotic orbit $x=0$, $y=-0.01597$,
$p_x\simeq0.49974$, $p_y=0$ of the 2D system (\ref{eq:2DHam}).}
\label{fig:2Dcomp}
\end{figure}

On the other hand, the computation of the maximal LCE, using
Eq.~(\ref{eq:lyap}), despite its usefulness in many cases, does
not have the same convergence properties over the same time
interval. This becomes evident in Fig. \ref{fig:2Dcomp}
where we plot the evolution of the $L_t$ (panel (a)) and  the SALI
(panel (b)) for a chaotic orbit of Eqs. (\ref{eq:2DHameq}). At $t
\approx 1900$ the SALI reaches the value $10^{-16}$ and no further
computations are needed. Of course we could be sure for the
chaotic nature of the orbit before that time, for example at
$t\approx 1000$ (where the $\mbox{SALI}\approx 10^{-8}$). On the
other hand, the computation of the $L_t$ (Fig.~\ref{fig:2Dcomp}(a))
up to $t\approx 1000$ or even up to $t\approx 1900$, still shows no
clear evidence of convergence. Although Fig.~\ref{fig:2Dcomp}(a)
suggests that the orbit might probably be chaotic, it does not allows
us to conclude its chaotic nature with certainty, and so further
computations of  $L_t$ are needed. Thus, it becomes evident that an
advantage of the SALI, with respect to the computation of $L_t$,
is that the current value of the SALI is sufficient to
determine the chaotic nature of an orbit, in contrast to the
maximal LCE where the whole evolution of the deviation vector
affects the computed value of $L_t$.

\section{The behavior of the SALI for ordered and chaotic motion}
\label{Behavior}

As we have seen for  ordered orbits the SALI remains different
from zero fluctuating around some non--zero value. The behavior of
the SALI for ordered motion was studied  and explained in detail
in the  case of a completely integrable 2D Hamiltonian
\cite{SABV03}, in which no chaotic orbits exist. It was shown that
any pair of arbitrary deviation vectors tend to the tangent space
of the torus, on which the motion is governed by 2 independent
vector fields, corresponding to the 2 integrals of motion. Thus,
since $\vec{w}_1(t)$ and $\vec{w}_2(t)$, in general have one
component `along' and one `across' the torus, there is no reason
why they should become aligned and thus typically end up
oscillating about two different directions. This  explains why the
SALI does not go to zero in the case of ordered motion.

Now let us investigate the dynamics in the vicinity of chaotic
orbits of a Hamiltonian system of $n$ degrees of freedom. An orbit
of this system is defined by $\vec{x}=(q_1, \, q_2,\, \ldots, \,
q_n, \,p_1, \, p_2,\, \ldots, \, p_n)$, with $q_i$, $p_i$,
$i=1,\ldots,n$ being the generalized coordinates and the conjugate
momenta respectively. The time evolution of this orbit is given by
 Hamilton's equations of motion
\begin{equation}
\frac{d \vec{x}}{dt}= \vec{V}(\vec{x})= \left( \frac{\partial
H}{\partial \vec{p}}\, ,  - \frac{\partial H}{\partial \vec{q}}
\right). \label{eq:Hameq}
\end{equation}
Solving the variational equations about a solution of
(\ref{eq:Hameq}), $\vec{x}(t)$, which represents our reference
orbit under investigation,
\begin{equation}
\frac{d \vec{w}}{dt} = M(\vec{x}(t)) \,\vec{w} \, , \label{eq:wt}
\end{equation}
where $M= \frac{\partial \vec{V}}{\partial \vec{x}}$ is the
Jacobian matrix of $\vec{V}$, we get the time evolution of an
initial deviation vector $\vec{w}(t_0)$, for sufficiently small
intervals $[t_0,t_0 + \Delta t]$. Note now that, in this context,
the eigenvalues $\lambda_1 \geqslant \lambda_2 \geqslant \cdots
\geqslant \lambda_{2n}$ of $M$, at $t=t_0$, may be thought of as
local Lyapunov exponents,  with $\hat{e}_1 , \, \hat{e}_2 , \,
\ldots, \, \hat{e}_{2n}$  the corresponding unitary eigenvectors.
These eigenvalues in fact oscillate about their time averaged
values, $\sigma_1 \geqslant \sigma_2 \geqslant \cdots \geqslant
\sigma_{2n}$, which are the global LCEs of the dynamics in that
region. As is well--known in Hamiltonian systems, the Lyapunov
exponents of chaotic orbits are real and are grouped in pairs of
opposite sign with at least two of them being equal to zero
\cite{LL92}. Thus, the evolution of any initial deviation vector
$\vec{w}_1 (0)$ is given by:
\begin{equation}
\vec{w}_1 (t) = \sum_{i=1}^{2n} c_i^{(1)}\, e^{\lambda_i t}\,
\hat{e}_i \, ,\label{eq:w1_1}
\end{equation}
where the $c_i^{(1)}$ are in general complex numbers and
$\lambda_i$, $\hat{e}_i$ depend on the specific location in phase
space, $\vec{x}(t_0)$, through which our orbit passes. Note that
we consider here only real eigenvalues and hence real
$c_i^{(1)}$. If some of the $\lambda_i$ are complex, their
corresponding contribution to (\ref{eq:w1_1}) will be oscillatory
and will not affect the argument that follows.

To get a first rough idea of the way the SALI evolves, we now make
some approximations on the evolution of this deviation vector.
First let us assume that the $\lambda_i$ do not fluctuate
significantly about their averaged values and hence can be
approximated by them, i.e.\ $\lambda_i \approx \sigma_i$.
Secondly, we consider that the major contribution to $\vec{w}_1
(t)$ comes from the two largest terms of Eq. (\ref{eq:w1_1}), so
that:
\begin{equation}
\vec{w}_1 (t) \approx c_1^{(1)}\, e^{\lambda_1 t}\, \hat{e}_1 \, +
c_2^{(1)}\, e^{\lambda_2 t}\, \hat{e}_2 \approx c_1^{(1)}\,
e^{\sigma_1 t}\, \hat{e}_1 \, + c_2^{(1)}\, e^{\sigma_2 t}\,
\hat{e}_2. \label{eq:w1_2}
\end{equation}
In this approximation, we now use (\ref{eq:w1_2}) to derive a
leading order estimate of the ratio
\begin{equation}
\frac{\vec{w}_1 (t)}{\| \vec{w}_1 (t)\|}\approx \frac{c_1^{(1)}\,
e^{\sigma_1 t}\, \hat{e}_1 \, + c_2^{(1)}\, e^{\sigma_2 t}\,
\hat{e}_2}{|c_1^{(1)}| \, e^{\sigma_1 t}} = s_1 \, \hat{e}_1 +
\frac{c_2^{(1)}}{|c_1^{(1)}|} \, e^{-(\sigma_1-\sigma_2)t}
\,\hat{e}_2\, , \label{eq:w1_4}
\end{equation}
and an entirely analogous expression for a second deviation
vector:
\begin{equation}
\frac{\vec{w}_2 (t)}{\| \vec{w}_2 (t)\|}\approx \frac{c_1^{(2)}\,
e^{\sigma_1 t}\, \hat{e}_1 \, + c_2^{(2)}\, e^{\sigma_2 t}\,
\hat{e}_2}{|c_1^{(2)}| \, e^{\sigma_1 t}} = s_2 \, \hat{e}_1 +
\frac{c_2^{(2)}}{|c_1^{(2)}|} \, e^{-(\sigma_1-\sigma_2)t}
\,\hat{e}_2 \, , \label{eq:w2_4}
\end{equation}
where $s_i= {\rm sign} (c_1^{(i)})$, $i=1,\,2$.

In order to compute the SALI, as defined by (\ref{eq:SALI}) we add
and subtract Eqs. (\ref{eq:w1_4}) and (\ref{eq:w2_4}) keeping the
norm of the minimum of the two evaluated quantities. Thus,
$\hat{e}_1$ does not appear in the expression of the SALI, which
becomes
\begin{equation}
\mbox{SALI}(t)= \min  \left\|
\frac{\vec{w}_1(t)}{\|\vec{w}_1(t)\|}\pm
\frac{\vec{w}_2(t)}{\|\vec{w}_2(t)\|} \right\| \approx \left|
\frac{c_2^{(1)}}{|c_1^{(1)}|}  \pm
\frac{c_2^{(2)}}{|c_1^{(2)}|}\right| \, e^{-(\sigma_1-\sigma_2)t}
\|\hat{e}_2\| \label{eq:saliap_1}
\end{equation}
Denoting by $c$ the positive  quantity on the r.h.s of the above
equation and using the fact that $\hat{e}_2$ is a unitary vector
we get
\begin{equation}
\mbox{SALI}(t) \approx c \, e^{-(\sigma_1-\sigma_2)t}\, .
\label{eq:saliap_2}
\end{equation}
Eq.~(\ref{eq:saliap_2}) clearly suggests that the SALI for chaotic
orbits tends to zero exponentially and  the rate of this decrease
is related to the two largest LCEs of the dynamics.

\begin{figure}
\centerline{\includegraphics[width=15 cm,height=7.5 cm]
{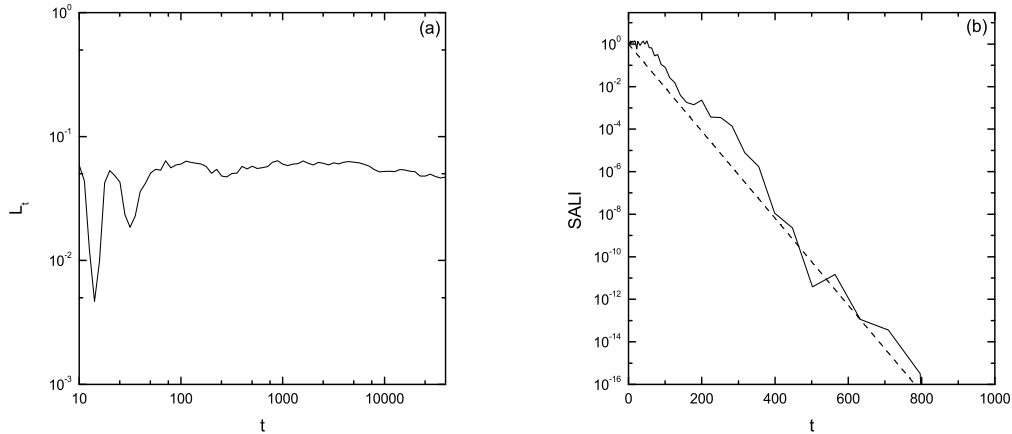}} \caption{(a) The evolution of $L_t$ for the chaotic
orbit with initial condition $x=0$, $y=-0.25$, $p_x\simeq0.42081$,
$p_y=0$ of the 2D system (\ref{eq:2DHam}). (b) The SALI of the
same orbit (solid line) and the function $e^{-\sigma_1 t}$ (dashed
line) for $\sigma_1=0.047$. Note that the t--axis is linear.}
\label{fig:2Dfit}
\end{figure}

Let us test the validity of this result by recalling  that 2D
Hamiltonian systems have only one positive LCE $\sigma_1$, since
the second largest is $\sigma_2=0$. So, Eq.~(\ref{eq:saliap_2})
becomes for such systems:
\begin{equation}
\mbox{SALI}(t) \approx c \, e^{-\sigma_1 t} \label{eq:saliap_2D}\,
.
\end{equation}
In Fig.~\ref{fig:2Dfit}(a)
we plot in log--log scale $L_t$  as a function of time $t$ for a
chaotic orbit of the 2D system (\ref{eq:2DHam}). $L_t$ remains
different from zero, which implies the chaotic nature of the
orbit. Following its evolution for a sufficiently long time
interval to obtain reliable estimates ($t \approx 10000$) we
obtain $\sigma_1 \approx0.047$. In Fig. \ref{fig:2Dfit}(b) we plot
the SALI for the same orbit (solid line) using linear scale for
the time $t$. Again we conclude that the orbit is chaotic as
$\mbox{SALI}\approx 10^{-16}$ for $t\approx800$. If Eq.
(\ref{eq:saliap_2D}) is valid, the slope of the SALI in Fig.
\ref{fig:2Dfit}(b) should be given approximately by $\sigma_1$,
being actually $- \sigma_1 / \ln 10$, because $\log(\mbox{SALI})$
is a linear function of $t$. As we are only interested in the
slope of the SALI, we plot in Fig. \ref{fig:2Dfit}(b) Eq.
(\ref{eq:saliap_2D}) for an appropriate value of $c$ (here $c=1$)
and $\sigma_1 = 0.047$ (dashed line) and find that  the agreement
of the approximate formula (\ref{eq:saliap_2}) to the computed
values of the SALI is indeed quite satisfactory.

\begin{figure}
\centerline{\includegraphics[width=15 cm,height=7.5 cm]
{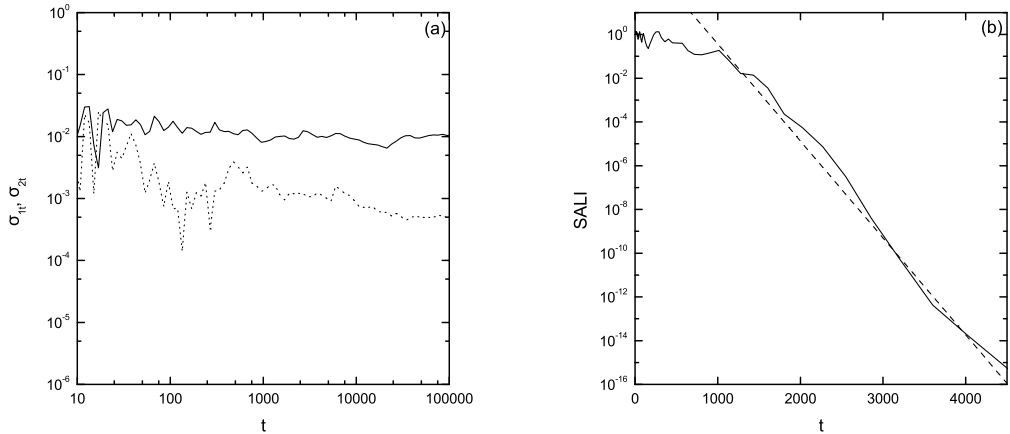}} \caption{(a) The evolution of the two Lyapunov
exponents $\sigma_{1t}$, $\sigma_{2t}$ for the chaotic orbit with
initial condition $x=0$, $y=0$, $z=0$, $p_x=0$, $p_y=0$,
$p_z\simeq 0.123693$ of the 3D system (\ref{eq:3DHam}). (b) The
SALI of the same orbit (solid line) and the function $c \,
e^{-(\sigma_1 - \sigma_2) t}$ (dashed line) for $\sigma_1 =
0.0107$, $\sigma_2 =0.0005$ and $c=10^5$. Note that the t--axis is
linear.} \label{fig:3Dfit}
\end{figure}

Chaotic orbits of 3D Hamiltonian systems generally have two
positive Lyapunov exponents,  $\sigma_1$ and $\sigma_2$. So, for
approximating the behavior of the SALI by Eq. (\ref{eq:saliap_2}),
both $\sigma_1$ and $\sigma_2$ are needed. We compute $\sigma_1$,
$\sigma_2$ for a chaotic orbit of the 3D system (\ref{eq:3DHam})
as the long time estimates of  some appropriate quantities
$\sigma_{1t}$, $\sigma_{2t}$ by applying the method proposed by
Benettin et al. \cite{BGGS80b}. The results are presented in
Fig.~\ref{fig:3Dfit}(a).
The computation is carried out until $\sigma_{1t}$ and
$\sigma_{2t}$ stop having high fluctuations and approach some
non--zero values (since the orbit is chaotic), which could be
considered as good approximations of their limits $\sigma_1$,
$\sigma_2$. Actually for $t \approx 10^5$ we have $\sigma_{1t}
\approx 0.0107$, $\sigma_{2t} \approx 0.0005$. Using these values
as good approximations of $\sigma_1$, $\sigma_2$ we see in Fig.
\ref{fig:3Dfit}(b) that the slope of the SALI (solid line) is well
reproduced by Eq. (\ref{eq:saliap_2}) (dashed line). Note how much
more quickly the SALI's convergence to zero shows chaotic
behavior, while the two LCEs, $\sigma_{1t}$ $\sigma_{2t}$, take a
lot longer to reach their limit values. Moreover, the results
presented in Figs. \ref{fig:2Dfit} and \ref{fig:3Dfit} give strong
evidence for the validity of Eq. (\ref{eq:saliap_2}) in describing
the behavior of the SALI for chaotic motion. So we conclude that
for chaotic motion the SALI is related to the two largest Lyapunov
exponents and decreases asymptotically as ${\mbox SALI}
\propto e^{-(\sigma_1-\sigma_2)t}$.

\section{Distinguishing between regions of order and chaos}
\label{Discussion}

The SALI offers indeed an easy and  efficient method for
distinguishing the chaotic vs.\ ordered nature of orbits in a
variety of problems. In the present section we use it for
identifying regions of phase space where large scale ordered and
chaotic motion are both present.

\begin{figure}
\centerline{\includegraphics[width=7.5 cm,height=7.5 cm]
{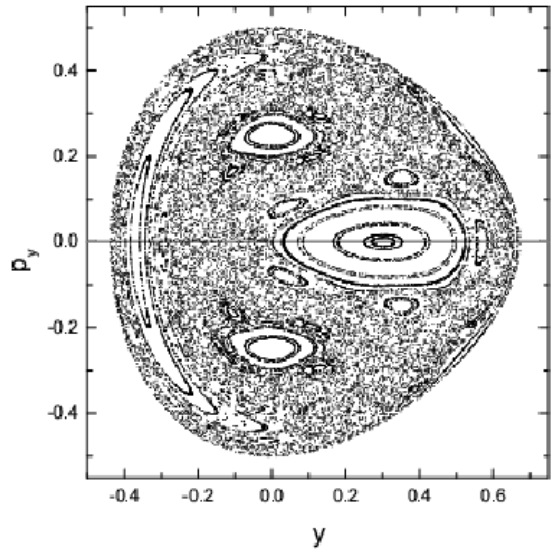}} \caption{The PSS $x=0$ of the 2D H\'{e}non--Heiles
system (\ref{eq:2DHam}). The axis $p_y=0$ is also plotted.}
\label{fig:2DPSS}
\end{figure}

\begin{figure}
\centerline{\includegraphics[width=15 cm,height=7.5 cm]
{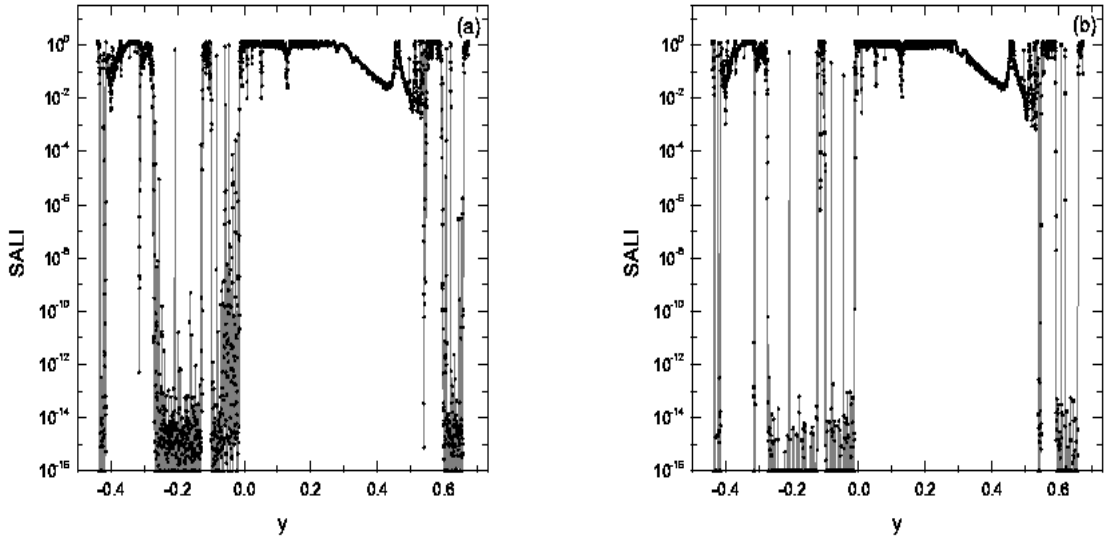}} \caption{The values of the SALI for (a) $t=1000$ and
(b) $t=4000$ for orbits of the 2D system (\ref{eq:2DHam}) with
initial conditions on the $p_y=0$ line on the PSS (Fig.
\ref{fig:2DPSS}), as a function of the $y$ coordinate of the
initial condition.} \label{fig:line}
\end{figure}

In Fig.~\ref{fig:2DPSS}
we present a  detailed plot of the $x=0$ PSS  of the 2D
H\'{e}non--Heiles system (\ref{eq:2DHam}). Regions of ordered
motion, around stable periodic orbits, are seen to  coexist with
chaotic regions filled by scattered points. In order to
demonstrate the effectiveness of the SALI method, we first
consider orbits whose initial conditions lie on the line $p_y=0$.
In particular we take $5000$ equally spaced initial conditions on
this line and compute the value of the SALI for each one. The
results are presented in Fig. \ref{fig:line}
where we plot the SALI as a function of the $y$ coordinate of the
initial condition of these orbits for $t=1000$ (panel (a)) and
$t=4000$ (panel (b)). In both panels the data points are line
connected, so that the changes of the SALI values are clearly
visible. Note that there are intervals  where the SALI has large
values (e.g. larger than $10^{-4}$), which correspond to ordered
motion in the island of stability crossed by the $p_y=0$ line in
Fig. \ref{fig:2DPSS}. There also exist regions where the SALI has
very small values (e.g. smaller than $10^{-12}$) denoting that in
these regions the motion is chaotic. These intervals correspond to
the regions of scattered points crossed by the $p_y=0$ line in
Fig. \ref{fig:2DPSS}. Although most of the initial conditions give
large ($> 10^{-4}$) or very small ($ \leqslant 10^{-12}$) values
for the SALI, there also exist initial conditions that have
intermediate values of the SALI ($10^{-12} < \mbox{SALI} \leqslant
10^{-4}$) e.g.  at $t=1000$ in Fig. \ref{fig:line}(a). These
initial conditions correspond to sticky chaotic orbits, remaining
for long time intervals at the borders of islands, whose chaotic
nature will be revealed later on.  By comparing Figs.
\ref{fig:line}(a) and \ref{fig:line}(b) it becomes evident that
almost all points having $10^{-12} < \mbox{SALI} \leqslant
10^{-4}$ in Fig. \ref{fig:line}(a)  move downwards to very small
values of the SALI in Fig. \ref{fig:line}(b), while the intervals
that correspond to ordered motion remain the same. Again in Fig.
\ref{fig:line}(b) there exist few points having intermediate
values of the SALI, which correspond to sticky orbits whose SALI
will eventually become zero. We note that it is not easy to
define a threshold value, so that the SALI being smaller than this value
reliably signifies chaoticity. Nevertheless, numerical
experiments in several systems show that in general a good guess
for this value could be $\lesssim 10^{-4}$.

In both panels of Fig. \ref{fig:line}, around $y\approx -0.1$
there exists a group of points inside a big chaotic region having
$\mbox{SALI}>10^{-4}$. These points correspond to orbits with
initial conditions inside a small stability island, which is not
even visible in the PSS of Fig. \ref{fig:2DPSS}. Also the point
with $y=-0.2088$  has very high value of the SALI ($> 0.1$) in
both panels of Fig. \ref{fig:line}, while all its neighboring
points have $\mbox{SALI}<10^{-9}$ even for $t=1000$. This point
actually corresponds to an ordered orbit inside a tiny island of
stability, which can be revealed only after a very high
magnification of this region of the PSS. So, we see that the
systematic application of the SALI method can reveal very fine
details of the dynamics.

\begin{figure}
\centerline{\includegraphics[width=7 cm,height=7 cm]{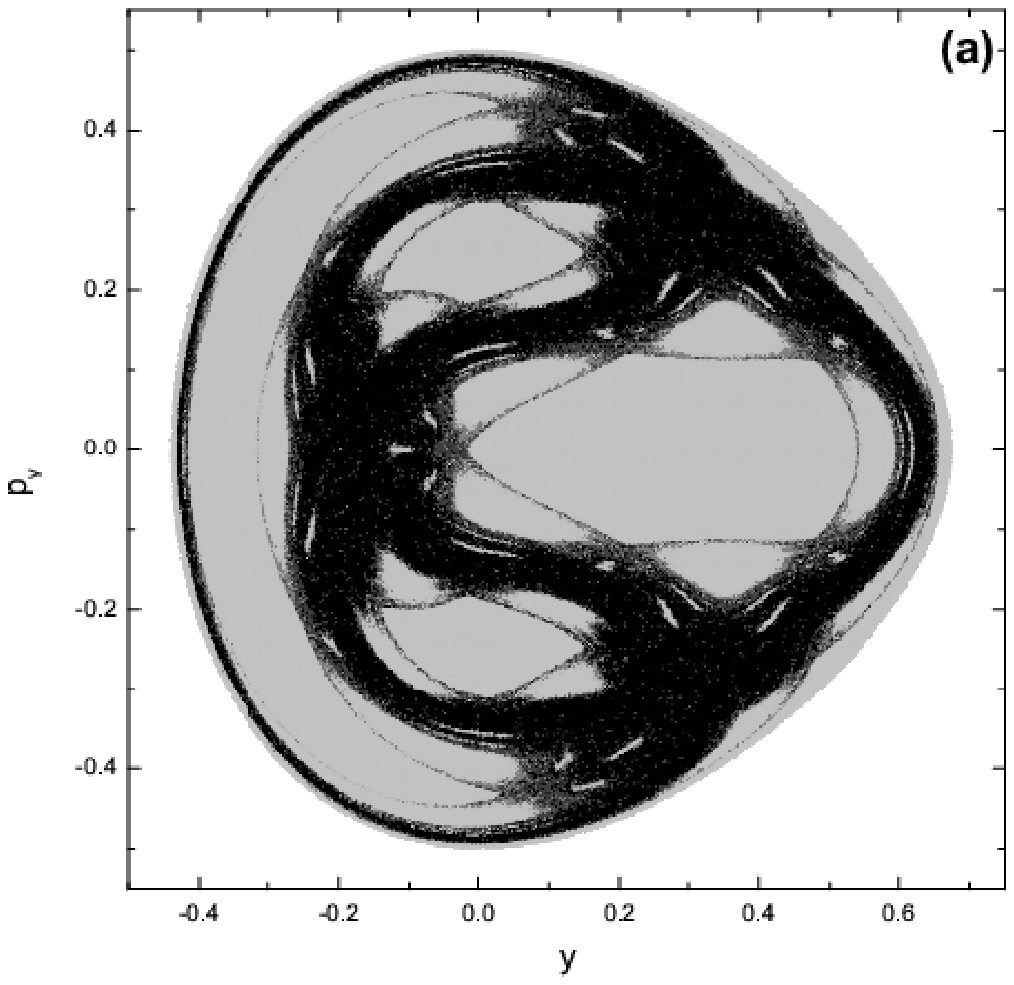}
\includegraphics[width=7 cm,height=7 cm] {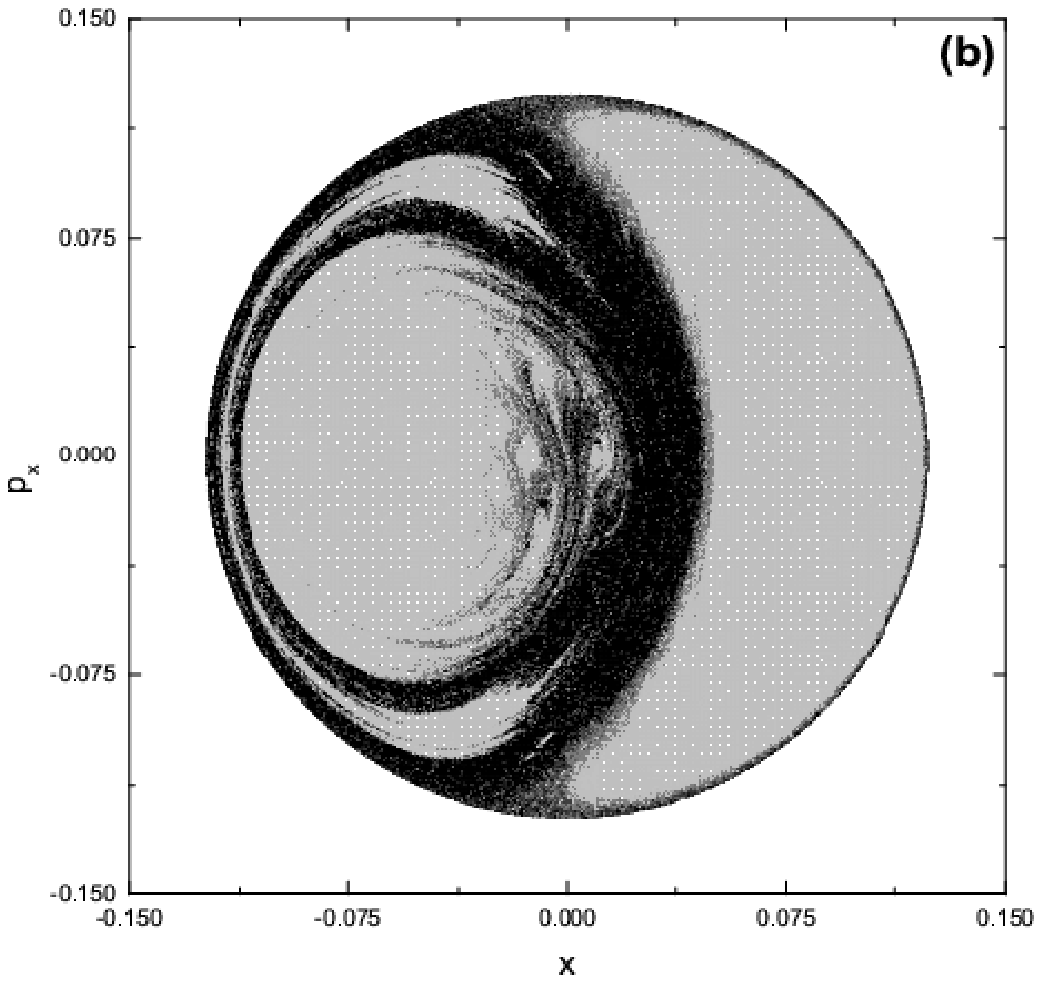} }
\caption{Regions of different values of the SALI (a) on the PSS
$x=0$ of the 2D H\'{e}non--Heiles system (\ref{eq:2DHam}), at
$t=1000$ and (b) on the subspace $y=0$, $p_y=0$ of the PSS $z=0$
of the 3D system (\ref{eq:3DHam}), at $t=5000$. In both frames
initial conditions are colored black if their $\mbox{SALI}
\leqslant 10^{-12}$, deep gray if $10^{-12} < \mbox{SALI}
\leqslant10^{-8}$, gray if $10^{-8} < \mbox{SALI} \leqslant
10^{-4}$ and light gray if $\mbox{SALI} > 10^{-4}$.}
\label{fig:2Dcolor}
\end{figure}

By carrying out the above analysis for points not only along a
line but on the whole plane of the PSS, and giving to each point a
color according to the value of the SALI, we can have a clear
picture of the regions where chaotic or ordered motion occurs. The
outcome of this procedure for the 2D H\'{e}non--Heiles system
(\ref{eq:2DHam}), using a dense grid of initial conditions on the
PSS, is presented in Fig.~\ref{fig:2Dcolor}(a).
The values of the logarithm of the SALI are divided in 4
intervals. Initial conditions having different values of the SALI
at $t=1000$ are plotted by different shades of gray: black if
$\mbox{SALI} \leqslant 10^{-12}$, deep gray if $10^{-12} <
\mbox{SALI} \leqslant 10^{-8}$, gray if $10^{-8} < \mbox{SALI}
\leqslant 10^{-4}$ and light gray if $\mbox{SALI} > 10^{-4}$.
Thus, in Fig. \ref{fig:2Dcolor}(a) we clearly distinguish between
light gray regions, where the motion is ordered and black regions,
where it is chaotic. At the borders between these  regions we find
deep gray and gray points, which correspond to sticky chaotic
orbits. It is worth--mentioning that in Fig.~\ref{fig:2Dcolor}(a)
we can see small islands of stability inside the large chaotic
sea, which are not visible in the PSS of Fig.~\ref{fig:2DPSS},
like the one for $y \approx -0.1$, $p_y \approx 0$. Although
Fig.~\ref{fig:2Dcolor}(a) was computed for only $t=1000$ (like
Fig.~\ref{fig:line}(a)), this time was sufficient for the clear
revelation of  small ordered regions inside the chaotic sea.

The construction of Fig.~\ref{fig:2Dcolor}(a) was actually speeded
up by attributing the final value of the SALI (at $t=1000$) of an
orbit to all  its intersection points with the PSS, and by
stopping the evolution of the orbit if its SALI became equal to
zero for $t<1000$. For a grid of $375\times 750$ equally spaced
initial conditions on the $p_y \geqslant 0$ part of the figure, we
need about 2 hours of CPU time on a Pentium 4 2GHz PC. Although it
is difficult to estimate, we expect that it would take
considerably longer to discern the same kind of detail, by
straightforward integration of Eqs.~(\ref{eq:2DHameq}) for a
similar grid of initial conditions.

For 3D Hamiltonians the PSS is 4--dimensional and thus, not so
useful as in the 2D case. On the other hand, the SALI can again
identify successfully regions of order and chaos in  phase space.
To see this, let us start with initial conditions on a 4d grid of
the PSS and attribute again the final value of the SALI of an
orbit to all the points visited by the orbit. In this way, we
again find regions of order and chaos, which may be visualized, if
we restrict our study to a subspace of the whole 4d phase space.
As an example, we plot  in Fig.~\ref{fig:2Dcolor}(b) the subspace
$y=0$, $p_y=0$ of the 4d PSS $z=0$ of the 3D system
(\ref{eq:3DHam}), using the same technique as in Fig.
\ref{fig:2Dcolor}(a). Again we can see regions of ordered (colored
in light gray) and chaotic motion (colored in black), as well as
sticky chaotic orbits (colored in deep gray and gray) at the edges
of these regions. Pictures like the ones of Fig.
\ref{fig:2Dcolor}, apart from presenting the regions of order and
chaos, could also be  used to estimate roughly the fraction of
phase space volume occupied by chaotic or ordered orbits and
provide good initial guesses for the location of stable periodic
orbits, in regions where the motion is ordered.

\section{Comparison with other methods}
\label{Compare}

The results presented in the previous sections show that the SALI
is a simple, efficient and easy to compute tool for distinguishing
between ordered and chaotic motion. Implementing the SALI is an
easy computational task as we only have to follow the evolution of
an orbit and of two deviation vectors, computing in every time
step the minimum norm of the difference and the addition of these
vectors. In the case of chaotic motion the SALI eventually tends
exponentially to zero, reaching rapidly very small values or even
the limit of the accuracy of the computer. On the other hand, in
the case of ordered motion the SALI fluctuates around non--zero
values. It is exactly this different behavior of the SALI that
makes it an ideal tool of chaos detection. The SALI has a clear
physical meaning as zero, or a very small value of the index,
signifies the alignment of the two deviation vectors. An advantage
of the  method is  that the index ranges in a defined interval
($\mbox{SALI} \in [0,\sqrt{2}]$) and so very small values of the
SALI (e.\ g.\ smaller than $10^{-8}$) establish the chaotic nature
of an orbit beyond any doubt.

The SALI helps us decide the chaotic nature of orbits faster and
with less computational effort than the estimation of the maximal
LCE. This happens  because the time needed for $L_t$ to give a
clear and undoubted indication of convergence to non--zero values
is usually much greater than the time in which the SALI becomes
practically zero, as can be seen in Figs. \ref{fig:2Dcomp},
\ref{fig:2Dfit} and \ref{fig:3Dfit}.

Many other chaos indicators have been introduced in recent years,
some of which are  compared in this section with the SALI. We also
study in  more detail the latest method we very recently became
aware of, the so--called 0--1 test, introduced by Gottwald and
Melbourne \cite{GM04}.

The efficiency of the SALI was compared in \cite{SALI} with the
well--known method of the Fast Lyapunov Indicator (FLI)
\cite{FLG97,FGL97} and the method of the spectral distance $D$ of
spectra of stretching numbers \cite{VCE99}. It was shown that the
SALI has comparable behavior to the FLI both for ordered and
chaotic orbits, with the SALI being able to decide the nature of
an orbit at least as fast as the FLI. An advantage of the SALI
method with respect to the FLI is the fact that the SALI ranges in
a given interval, with very small values corresponding to chaotic
behavior, while the values of FLI increase in time, both for
ordered and chaotic motion, but with different rates. So, the
interpretation of different colors in color plots produced by the
SALI method, like the ones of Fig.\ \ref{fig:2Dcolor}, does not
depend on the integration time of the orbits, in contrast to
similar plots of the FLI, since the range of FLI values changes as
time grows. As   was explained in detail in \cite{SALI} the
computation of the SALI is much easier and faster than the
computation of the spectral distance $D$, mainly because we do not
have to go through the computation of the spectra of stretching
numbers. Also the SALI can be used to distinguish between order
and chaos  in the case of 2d maps, where the spectral distance $D$
cannot be applied.

S\'{a}ndor et al. \cite{SESF04} comparing the Relative Lyapunov
Indicator (RLI) method with the SALI  showed that both indices
have similar behaviors even in cases of weakly chaotic motion. The
RLI is practically the absolute value of the difference of $L_t$
of two initially nearby orbits, with very small values of the RLI
denoting ordered motion, while large differences between the
$L_t$'s denote chaotic behavior (for more information on the RLI
method see \cite{SESF04}). We note that the computation of the RLI
requires the time evolution of two orbits and two deviation
vectors (one for each orbit), while the computation of the SALI is
faster as we compute one orbit and two deviation vectors.

Very recently, Gottwald and Melbourne \cite{GM04} introduced a new
test for distinguishing  ordered from chaotic behavior in
deterministic dynamical systems: the 0--1 test. The method is
quite general and can be applied directly to long time series data
produced by the evolution of a dynamical system. In that sense,
the 0--1 test is more general than the SALI for the computation of
which we need to know the equations that govern the evolution of
the system, as well as its variational equations. The 0--1 test
uses the real valued function:
\begin{equation}
p(t)= \int_0^t \phi(s) \, \cos(\psi(s)) \, ds \,\, ,\label{01p}
\end{equation}
where $\phi(s)$ is, in general, any observable of the underlying
dynamics and
\begin{equation}
\psi(t)=kt+ \int_0^t \phi(s) \, ds \,\, , \label{01psi}
\end{equation}
with $k$ being a positive constant. By defining the
mean--square--displacement of p(t):
\begin{equation}
M(t)=\lim_{T\rightarrow \infty}  \frac{1}{T} \int_0^T \left(
p(t+\tau)-p(\tau)\right)^2\, d\tau \,\, , \label{01M}
\end{equation}
and setting
\begin{equation}
K(t)= \frac{\log (M(t)+1)}{\log t}\, , \label{01Kt}
\end{equation}
one takes the limit
\begin{equation}
K= \lim_{t\rightarrow \infty} K(t)\, ,\label{01K}
\end{equation}
and characterizes the particular orbit as ordered if $K=0$, or
chaotic if $K=1$. The justification of the 0--1 test, as well as
applications of the method to some dynamical systems can be found
in \cite{GM04}.

\begin{figure}
\centerline{\includegraphics[width=7.5 cm,height=7.5 cm]
{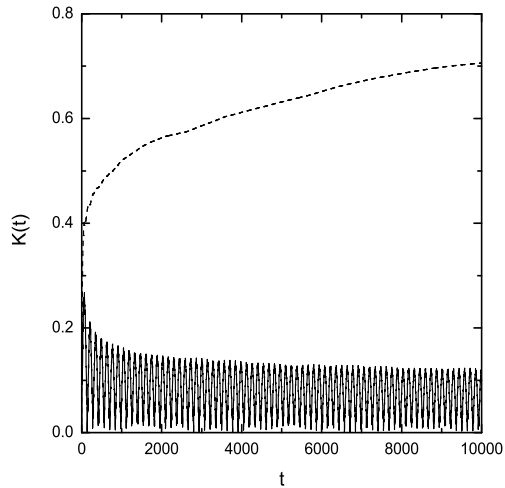}} \caption{Application of the 0--1 test for the orbits
of Fig.\ \ref{fig:2Dexample}. $K(t)$ oscillates with an amplitude
decreasing towards  $K=0$ for the ordered orbit (solid line),
while it tends to $K=1$ for the chaotic orbit (dashed line). }
\label{fig:01}
\end{figure}

In order to compare the 0--1 test with the SALI method we apply it
in the cases of the ordered and chaotic orbits of the
H\'{e}non--Heiles system (\ref{eq:2DHam}) presented in Fig.\
\ref{fig:2Dexample}. Recall that in the case of the chaotic orbit
the SALI determines the true nature of the orbit at $t\approx 800$
when $\mbox{SALI} \approx 10^{-16}$, or even at $t\approx 400$ if
we consider the more loose condition that $\mbox{SALI} \approx
10^{-8}$ guarantees chaoticity. We consider as an observable
$\phi(t)$ the quantity:
\begin{equation}
\phi(t)=y(t)+p_y(t)\, , \label{01HH}
\end{equation}
while for the constant $k$ we adopt the value used in \cite{GM04},
i.\ e.\ $k=1.7$. The application of the 0--1 test requires a
rather long time series of the observable $\phi(t)$ in order to
reliably compute firstly $M(t)$ for $T\rightarrow \infty $ (Eq.\
\ref{01M}) and secondly $K$ as the limit of $K(t)$ for
$t\rightarrow \infty$ (Eq.\ \ref{01K}). In our computations we set
$T=90000$ time units and compute $K(t)$ for $t\in (0, 10000)$,
which means that the particular orbit is integrated up to $t=10^5$
time units. Although the assumption  $T\gg t$, which is formally
required for the convergence of $K(t)$ (see \cite{GM04}), is not
fulfilled, the behavior of $K(t)$ for the ordered and chaotic
orbit is different, allowing us to distinguish between the two
cases (Fig.\ \ref{fig:01}).
In particular, $K(t)$ increases for the chaotic orbit (dashed line
in Fig.\ \ref{fig:01}) showing a tendency to reach $K=1$, while it
tends to $K=0$ for the ordered orbit (solid line in Fig.\
\ref{fig:01}) exhibiting fluctuations of slowly decreasing
amplitude. From these results it is obvious that the true nature
of the orbits is determined correctly by the 0--1 test, but the
computational effort needed in order to be able to characterize
the orbits is much higher than in the case of the SALI. This
difference is due to, firstly,  the more complicated way of
estimating $K$, in comparison with the computation of the SALI, as
we have to compute several integrals, and secondly, the fact that
we must compute the particular orbit for sufficiently long time,
in order to approximate  the limits of Eqs.\ (\ref{01M}),
(\ref{01K}). Of course, as we have already mentioned, the 0--1
test is  more general  as it can also be readily applied to time
series data, without knowing necessarily the equations of the
dynamical system.

\section{Summary}
\label{Summary}

In this paper we have applied the SALI method to distinguish
between order and chaos in  2D and  3D autonomous Hamiltonian
systems, and have also analyzed the behavior of the index for
chaotic  orbits. Our results  can be summarized as follows.

\begin{itemize}

\item The SALI proves to be an ideal indicator of chaoticity
independent of the dimensions of the system. It tends to zero for
chaotic orbits, while it exhibits small fluctuations around
non--zero values for ordered ones and so it clearly distinguishes
between these two cases. Its  advantages are its simplicity,
efficiency and reliability as it can rapidly and accurately
determine the chaotic vs ordered nature of a given orbit. In
regions of `stickiness', of course, along the borders of ordered
motion it displays transient oscillations. However, once the orbit
enters a large chaotic domain the SALI converges exponentially to zero,
often at  shorter times than it takes the maximal Lyapunov
exponent to converge to its limiting value.

\item We emphasize that the main advantage of the SALI in chaotic
regions is that it uses two deviation vectors  and exploits at
every step, their convergence to the unstable manifold from all
previous steps. This allows us to show that the SALI tends to zero
for chaotic orbits at a rate which is related to the difference of
the two largest Lyapunov characteristic exponents $\sigma_1$,
$\sigma_2$ as $\mbox{SALI} \propto e^{-(\sigma_1-\sigma_2)t}$. By
comparison, the computation of the maximal LCE, even though it
requires only one deviation vector and one exponent, $\sigma_1$,
often takes longer to converge, since it needs to average over
many time intervals, where the calculation of this exponent is
independent from all previous intervals. The SALI was also proved
to have similar or even better performance than other methods of
chaos detection which were briefly discussed in Sec.\
\ref{Compare}.

\item The $\mbox{SALI} \in [0,\sqrt{2}]$ and its value
characterize an orbit of being chaotic or ordered. Exploiting this
feature of the index  we have plotted detailed  phase space
portraits both for 2D and 3D Hamiltonian systems, where the
chaotic and ordered regions are clearly distinguished. We were
thus able to trace in a fast and systematic way very small islands
of ordered motion, whose detection by traditional methods would be
very difficult and time consuming. This approach is therefore
expected to provide useful tools for the location of stable
periodic orbits, or the computation of the phase space volume
occupied by ordered or chaotic motion in multidimensional systems,
where the PSS is not easily visualized, and very few other similar
techniques of practical value are available.
\end{itemize}

\ack We acknowledge fruitful discussions on the contents of this
work  with Professors Giulio Casati and Tomas Prosen. We would
also  like to thank the anonymous referee for very useful comments
which helped us improve the clarity of the paper. This research
was partially supported by the `Heraclitus' research program of
the Greek Ministry of Development. Ch.\ Skokos was supported by
the `Karatheodory' post--doctoral fellowship No 2794 of the
University of Patras and Ch.\ Antonopoulos was supported by the
`Karatheodory' graduate student fellowship No 2464 of the
University of Patras.

\end{document}